\documentclass[twocolumn]{aa}
\usepackage{graphicx}
\usepackage{natbib}
\begin{document}

\title{A new definition of the intermediate group of
gamma-ray bursts}

 \author{I. Horv\'ath \inst{1}
\and
 L.G. Bal\'azs \inst{2}
 \and
Z. Bagoly \inst{3}
\and
F. Ryde \inst{4}
\and
A. M\'esz\'aros \inst{4,5,6} }

\offprints{I. Horv\'ath}

\institute{Department of Physics, Bolyai Military University, Budapest,
Box-12, H-1456, Hungary\\
           \email{horvath.istvan@zmne.hu}
   \and
              Konkoly Observatory, Budapest, Box-67, H-1525, Hungary\\
    \email{balazs@konkoly.hu}
   \and
              Laboratory for Information Technology, E\" otv\" os
          University, Budapest, P\'azm\'any P. s. 1/A,, H-1117,
          Hungary\\
          \email{bagoly@ludens.elte.hu}
  \and
           Stockholm Observatory, AlbaNova, SE-106 91 Stockholm, Sweden\\
            \email{felix@astro.su.se}
\and
               Astronomical Institute of the Charles University,
              V Hole\v{s}ovi\v{c}k\'ach 2, CZ 180 00 Prague 8,
          Czech Republic\\
              \email{meszaros@mbox.cesnet.cz}
\and
Max Planck Institute for Astrophysics, Garching,
   Karl-Schwarzschild-Str. 1, Postfach 1317, D-85741 Garching, Germany
}
   \date{Received April 20, 2004; accepted  September 25, 2005}

\abstract{Gamma-ray bursts can be divided 
into
three groups ("short", "intermediate", "long") with respect to
their durations. This classification is somewhat imprecise, since
the subgroup of intermediate duration has an admixture of both
short and long bursts. In this paper a physically more reasonable
definition of the intermediate group is presented, using also the
hardnesses of the bursts. It is shown again that the existence of
the three groups is real, no further groups are
needed. The intermediate group is
the softest one. From this new definition it follows that $11\%$
of all bursts belong to this group.
 An anticorrelation between the
hardness and the duration is found for this subclass in contrast
to the short and long groups. Despite this difference it is not
clear yet whether this group represents a physically different
phenomenon.
}

\authorrunning{Horv\'ath et al.}
\titlerunning{A new definition ...}
 \maketitle

\section{Introduction}

It is a great challenge to classify  gamma-ray bursts (GRBs) into
classes. Mazets et al. (1981) and Norris et al. (1984) 
suggested there might be a separation in the duration
distribution. Using the First BATSE Catalog, Kouveliotou et al.
(1993) found a bimodality in the distribution of the logarithms of
the durations. This bimodality is highly pronounced, if one uses
the parameter $T_{90}$ (the time during which 90\% of the fluence
is accumulated \cite{kou}) to characterize the durations of
GRBs \cite{mcb94,kos,belli,pen97}. Today it is widely accepted
that the physics of these two groups (also called "subclasses" or
simply "classes") are different, and these two kinds of
GRBs are different phenomena \cite{nor01,bal03}. 
The high redshifts and the cosmological distances are directly
confirmed for the long bursts only, while for the short ones there
is only indirect evidence for their cosmological origin
\cite{me01,me03}.

Using the Third BATSE Catalog \cite{m6} Horv\'ath (1998) has shown
that the distribution of the logarithms of the durations of GRBs
($\log T_{90}$) could be well fitted by a sum of three Gaussian
distributions. He finds it statistically unlikely (with a
probability $\sim 10^{-4}$) that there are only two groups.
Simultaneously Mukherjee et al. (1998) report the finding (in a
multidimensional parameter space) of a very similar group
structure of GRBs. Somewhat later several authors
\cite{hak,bala,rm,hak03,bor04,hak04} included more physical
parameters into the analysis of the bursts (e.g. peak-fluxes,
fluences, hardness ratios, etc.). A cluster analysis in this
multidimensional parameter space suggests the existence of the
third ("intermediate") group as well \cite{muk98,hak,bala,rm}. The
physical existence of the third group is, however, still not
convincingly proven. For example, Hakkila et al. (2000) believe
that the third group is only a deviation caused by a complicated
instrumental effect, which can reduce the durations of some faint
long bursts. Later Hakkila et al. (2003) published another paper
which had different conclusions (we discuss this 
 greater detail later). However, the celestial distribution of the
third group is anisotropic \cite{bal98,bal99,me00a,me00b,li01};
i.e. different from that of the long GRBs alone \cite{mesto}. The
logN-logS distribution may also differ from those of the other
groups \cite{ho98}. Taken together this means that the existence
of the third intermediate group is acceptable, but its physical
meaning, importance and origin is less clear than those of the
other groups. Hence, its further study is required.

Using Principal Component Analysis (PCA), Bagoly et al. (1998) have
shown that there are only two major quantities necessary (called
the Principal Components; PCs) to characterize most of  the
properties of the bursts in the BATSE Catalog.  Consequently, the
problem of the choice of the relevant parameters describing GRBs
is basically a two-dimensional problem. For the statistical
analysis the choice of two independent parameters is enough; they
may be, but are not necessarily, the two principal components.
This means that only two parameters, relevantly chosen, should be
enough for the classification and determination of the groups.
Concluding from the analysis of the clustering properties of
GRBs in the BATSE 3B Catalog Mukherjee et al. (1998) identified
the following measured quantities relevant for classification:
duration ($T_{90}$), total fluence ($F_{tot}=F_1+F_2+F_3+F_4$) and
hardness ($H_{321}=\frac{F_3}{F_1+F_2}$). ($\log H_{321}$ 
is highly redundant with $\log H_{32}$ (= $\log F_3 - \log F_2$)
which is a linear combination of the two PCs mentioned above.)

In order to perform a statistical analysis to estimate the
probable number of classes Mukherjee et al. (1998) made the
apriori assumption that the observed BATSE sample is a
superposition of multivariate Gaussians in the variables included
in the analysis. Concerning $\log T_{90}$ Horv\'ath (1998) showed
that its distribution could be well fitted with three Gaussians.
Recently, Bal\'azs et al. (2003) has proven that the intrinsic
distributions of the total fluence and duration were  two
dimensional Gaussians  for the long and short GRBs, separately.
The Gaussian fit for the observed distribution of the total
fluence of long bursts, however, was poor due to the effect of the
luminosity distance. The dependence of the observed fluence
distribution on the luminosity distance might result in 'ghost
clusters' when  attempting to fit with Gaussians. In the contrary,
the effect of the luminosity distance was eliminated when
computing hardness.

Fitting the observed distribution with the superposition of
Gaussian components one had to keep the number of estimated
parameters as small as possible to ensure the stability of the
Maximum Likelihood procedure (e.g. in case of two dimensions and 4
components the number of parameters is 23 while the same in 3
dimensions is 39). Summarizing all these considerations we
decided to use two dimensional Gaussians with the logarithmic
duration ($\log T_{90}$) and hardness ($\log H_{321}$ or $\log
H_{32}$, alternatively).

Based on this technique several questions should be answered
concerning the intermediate group. First, will the statistical
analysis, using only these two parameters, reconfirm the existence
of the intermediate group? Second, if this question is answered in
the affirmative, then one has to show that either further groups
exist, or they do not. Using a much smaller sample Mukherjee
et al. (1998) claim that only three groups are necessary. On the
other hand, Cline et al. (1999) propose the existence of a fourth
subgroup of very short durations. Third, one also has to define
the quantities by which this third group is different. Fourth,
the method - making it possible to assign a certain GRB to a given
group - should also be developed. Fifth, the fraction of this
third intermediate group in the whole BATSE Catalog should also be
determined more exactly. Sixth, does the intermediate group really
represent a third type of bursts different from both the short and
long ones in its astrophysical origin?

The observational data from The Current BATSE GRB Catalog
\cite{mee01} will be used to answer these questions
in which there are 2702 GRBs, for 1956
of which both the  hardnesses and durations are measured.  The paper is
organized as follows: Section 2 briefly summarizes the mathematics
of the two-dimensional fits. Section 3 deals with these fits in
the two-dimensional parameter space and confirms the
reality of the intermediate group. Section 4 gives the
mathematical definition of the intermediate group making it
possible to determine, for any GRB, the  probability that it
belongs to a given group and deals with possible
observational bias. Section 5 discusses the physical differences
between the classes. Section 6 summarizes the conclusions of
this paper.

\section{Mathematics of the two-dimensional fit of $k$ classes}

We will study the distribution of GRBs in the $\{\log T_{90}; \log
H_{32}\}$ plane. Previously, Belli (1997) used this plane to
separate the bursts. She suggested that the curve $ H_{32} = 2
T_{90}^{0.5}$ gave a better division than the cut $ T_{90} = 2$ s
between the short and long GRBs.

We can assume that the observed probability distribution of the
GRBs in this plane is a superposition of the distributions
characterizing the different types of bursts present in the
sample. Introducing the notations $x= \log T_{90}$ and $y=\log
H_{32}$ and using the law of full probabilities \cite{renyi} we
can write
\begin{equation}\label{lfpr}
    p(x,y)
    =\sum \limits_{l=1}^k p(x,y|l)p_l.
\end{equation}
In this equation $p(x,y|l)$ is the conditional probability density
assuming that a burst belongs to the $l$-th class. $p_l$ is the
probability for this class in the observed sample ($\sum
\limits_{l=1}^k p_l = 1$), where $k$ is the number of classes. In
order to decompose the observed probability distribution $p(x,y)$
into the superposition of different classes we need the functional
form of $p(x,y|l)$. The probability distribution of the logarithm
of durations can be well fitted by Gaussian distributions, if we
restrict ourselves to the short and long GRBs \cite{ho98}. We
assume the same also for the $y$ coordinate. With this assumption
we obtain, for a certain $l$-th class of GRBs,
$$
p(x,y|l)  =
 \frac{1}{2 \pi \sigma_x \sigma_y
\sqrt{1-r^2}} \times \;\;\;\;\;\;\;\;\;\;\;\;\;\;\;\;\;\;\;\;
\;\;\;\;\;\;\;\;\;\;\;\;$$
\begin{equation} \label{gauss}
\exp\left[-\frac{1}{2(1-r^2)}
\left(\frac{(x-a_x)^2}{\sigma_x^2} + \frac{(y-a_y)^2}{\sigma_y^2}
- \frac{C} {\sigma_x \sigma_y}\right)\right], \;
\end{equation}
where $C = 2r(x-a_x)(y-a_y)$; $a_x$, $a_y$ are the means,
$\sigma_x$, $\sigma_y$ are the dispersions, and $r$ is the
correlation coefficient (\cite{tw53}; Chapt. 1.25). Hence, a
certain class is defined by 5 independent parameters, $a_x$,
$a_y$, $\sigma_x$, $\sigma_y$, $r$, which are different for
different $l$. If we have $k$ classes, then we have $(6k - 1)$
independent parameters (constants), because any class is given by
the five parameters of Eq.(\ref{gauss}) and the weight $p_l$ of
the class. One weight is not independent, because it holds $\sum
\limits_{l=1}^{k} p_l = 1$. The sum of $k$ functions defined by
Eq.(\ref{gauss}) gives the theoretical function of the fit. In
Bal\'azs et al. (2003) this fit for $k=2$ was used, and the
procedure for $k=2$ was described in more detail. However, that
paper used fluence instead of hardness. Here we will make similar
calculations  for $k=3$ and $k=4$.

\section{New confirmation of the intermediate group}

By decomposing $p(x,y)$
into the superposition of $p(x,y|l)$ conditional probabilities one
divides the original population of GRBs into $k$ groups, at least
from the mathematical point of view. Decomposing the left-hand side
of Eq.(\ref{lfpr}) into the sum of the right-hand side, one needs
the functional form of $p(x,y|l)$ distributions, and also $k$ has
to be fixed. Because we assume that the functional form is a
bivariate Gaussian distribution (see Eq.(\ref{gauss})), our task
is reduced to evaluate its parameters, $k$ and $p_l$.

In order to find the unknown constants in Eq.(\ref{gauss}) we use
the Maximum Likelihood (ML) procedure of parameter estimation
\cite{bal03}. Assuming a set of $N$ observed $[x_i, y_i], \,
(i=1, \dots, N)$ values ($N$ is the number of GRBs in the sample
for our case, which here is 1956) we can define the Likelihood
Function in the usual way, after fixing the value of $k$, in the
form
\begin{equation}\label{ml}
  L=\sum \limits_{i=1}^N \log p(x_i,y_i)\;,
\end{equation}
where $p(x_i,y_i)$ has the form given by Eq.(\ref{lfpr}).
Similarly, as it was done by Bal\'azs et al. (2003), the EM
(Expectation and Maximization) algorithm is used to obtain the
$a_x, a_y, \sigma_x, \sigma_y, r$ and $p_l$ parameters at which
$L$ reaches its maximum value. We made the calculations for
different values of $k$ in order to see the improvement of $L$ as
we increase the number of parameters to be estimated.

Tables \ref{tab1}-\ref{tab3} summarize the results of the fits for
$k=2,3,4$.

\begin{table}  \centering  \caption{Results of the EM algorithm in the $\{\log T_{90}; \log
H_{32}\}$ plane.
  $k=2$ \ \ $L_{max}=1193$  }
\label{tab1}
  \begin{tabular}{ccccccc} \hline
   $l $ & $p_l$  &  $a_x $ &  $a_y$ & $\sigma_x$ & $\sigma_y$  &  $r$ \\
 \hline
   1    &   .280  &  -.233  &   .740  &   .541  &   .259  &   .049 \\
  2    &   .720  &  1.488  &   .396  &   .471  &   .237  &   .128 \\
 \hline
\end{tabular}
\end{table}
\begin{table}
  \centering
  \caption{Results of the EM algorithm.  $k=3$ \ \ $L_{max}=1237$
  }\label{tab2}
  \begin{tabular}{ccccccc} \hline
  $l $ & $p_l$  &  $a_x $ &  $a_y$ & $\sigma_x$ & $\sigma_y$  &  $r$ \\
\hline
   1   &    .245  &  -.301  &   .763  &   .525  &   .251  &   .163 \\
   2   &    .109  &   .637  &   .269  &   .474  &   .344  &  -.513 \\
   3   &    .646  &  1.565  &   .427  &   .416  &   .210  &  -.034 \\
   \\ \hline
\end{tabular}
\end{table}

\begin{table}  \centering  \caption{Results of the EM algorithm.  $k=4$ \ \
 $L_{max}=1243$  }\label{tab3}
  \begin{tabular}{ccccccc} \hline
$l $ & $p_l$  &  $a_x $ &  $a_y$ & $\sigma_x$ & $\sigma_y$  & $r$
\\ \hline
 1   &    .234  &  -.307  &   .752  &   .524  &   .246  &   .215 \\
   2   &    .060  &   .441  &   .426  &   .637  &   .440  &  -.871 \\
   3   &    .060  &   .623  &   .262  &   .325  &   .325  &  -.095 \\
   4   &    .646  &  1.569  &   .426  &   .410  &   .211  &  -.034 \\
   \hline
\end{tabular}
\end{table}

The confidence interval of the parameters estimated can be given
on the basis of the following theorem. Denoting by $L_{max}$ and
$L_0$ the values of the Likelihood Function at the maximum and at
the true value of the parameters, respectively, one can write
asymptotically as the sample size $N\rightarrow\infty$
\cite{Kendall76},
\begin{equation}\label{lmax}
    2(L_{max}-L_0)\approx \chi_m^2,
\end{equation}
where $m$ is the number of parameters estimated ($m=6k -1$ in our
case), and $\chi_m^2$ is the usual $m$-dimensional $\chi^2$
function \cite{tw53}. Moving from $k=2$ to $k=3$ the number of
parameters $m$  increases by 6
 (from 11 to 17), and $L_{max}$  grows from
1193 to 1237. Since $\chi_{17}^2=\chi_{11}^2+\chi_6^2$ the
increase in $L_{max}$ by a value of 44 corresponds to a value of
88 for a $\chi_6^2$ distribution. The probability for $\chi_6^2
\geq 88$ is extremely low ($<10^{-10}$), so we may conclude that
the inclusion of a third class into the fitting procedure is well
justified by a very high level of significance.

Moving from $k=3$ to $k=4$, however, the improvement in $L_{max}$
is only 6 (from 1137 to 1143) corresponding to $\chi_6^2 \geq 12$,
which can happen by chance with a probability of 6.2 \%. Hence,
the inclusion of the fourth class is {\it not} justified. We may
conclude from this analysis that the superposition of three
Gaussian bivariate distributions - and {\it only these three ones} -
can describe the observed distribution.

This means that the 17 independent constants for $k=3$ in Table
\ref{tab2} define the parameters of the three groups. We see that
the mean hardness of the intermediate class is very low - the
third class is the softest one. Because $p_2 = 0.109$, 11\% of
all GRBs belongs to this group. This value is very close to those
found previously \cite{muk98,ho98,hak,ho02,rm,ho04}.

To test the robustness of the groups found by using this procedure
we also repeated the calculations in the $\{\log T_{90}; \log
H_{321}\}$ plane. Comparing the maximum  values of the likelihood
function (920, 980, 982)
obtained by assuming $k=$ 2, 3 and 4 components it is
clear from Tables \ref{tab22}, \ref{tab32} and \ref{tab42} that 3
Gaussian distributions are necessary and sufficient to account for the GRB
sample studied ($L_{max}^3-L_{max}^2=60$ , $L_{max}^4-L_{max}^3=2$).

\begin{table} \centering
\caption{Results of the EM algorithm in the $\{\log T_{90}; \log
H_{321}\}$ plane. $k=2$ \ \ $L_{max}=920$} \label{tab22}
\begin{tabular}{ccccccc}
\hline
$l $ & $p_l$  &  $a_x $ &  $a_y$ & $\sigma_x$ & $\sigma_y$  &  $r$ \\
\hline
1    &  0.276 &  -0.251 &   0.544 &   0.531 &   0.256 &   0.016 \\
2    &  0.725 &   1.479 &   0.132 &   0.479 &   0.287 &   0.123 \\
\hline
\end{tabular}
\end{table}

\begin{table} \centering \caption{Results of the EM algorithm.
$k=3$ \ \ $L_{max}=980$} \label{tab32}
\begin{tabular}{ccccccc}
\hline
$l $ & $p_l$  &  $a_x $ &  $a_y$ & $\sigma_x$ & $\sigma_y$  &  $r$ \\
\hline
1   &   0.233 &  -0.354 &   0.560 &   0.486  &  0.237 &   0.082 \\
2   &   0.154 &   0.722 &   0.057 &   0.480  &  0.432 &  -0.356 \\
3   &   0.613 &   1.588 &   0.174 &   0.404  &  0.249 &  -0.048 \\
\hline
\end{tabular}
\end{table}

\begin{table} \centering \caption{Results of the EM algorithm.
$k=4$ \ \ $L_{max}=982$} \label{tab42}
\begin{tabular}{ccccccc}
\hline
$l $ & $p_l$  &  $a_x $ &  $a_y$ & $\sigma_x$ & $\sigma_y$  &  $r$ \\
\hline
1  &    0.234 &  -0.354 &  0.559 &   0.485  &  0.238  &  0.078 \\
2  &    0.148 &   0.704 &  0.062 &   0.447  &  0.432  & -0.335 \\
3  &    0.333 &   1.580 &  0.115 &   0.403  &  0.268  & -0.141 \\
4  &    0.284 &   1.600 &  0.236 &   0.400  &  0.214  &  0.064 \\
\hline
\end{tabular}
\end{table}

Comparison of the results obtained in the $\{\log T_{90}; \log
H_{32}\}$ and $\{\log T_{90}; \log H_{321}\}$ planes show that the
parameters of the Gaussian distributions match each other well in the $x=\log
T_{90}$ coordinate (see Tables \ref{tab2} and \ref{tab32}).

\section{Mathematical classification of GRBs}

\subsection{The method}

Based on the calculations in the previous paragraph we resolved
the $p(x,y)$ probability density of the observed quantities into a
superposition of three Gaussian distributions. Using this
decomposition we can classify {\it any} observed GRB into the
classes represented by these groups. In other words, we develop a
method allowing us to obtain, for any given GRB, its three
membership probabilities, which  define the likelihood of the GRB
to  belong to the short, intermediate and long groups. The sum of
these three probabilities is unity. For this purpose we
define the following $I_l(x,y)$ indicator function, which assigns
to each observed burst a membership probability in a given $l$
class as follows:
\begin{equation}\label{mprob}
    I_l(x,y)= \frac{p_l p(x,y|l)}{\sum \limits_{l=1}^k p_l p(x,y|l)}.
\end{equation}

\begin{figure}
 \includegraphics[width=6.1cm, angle=270]{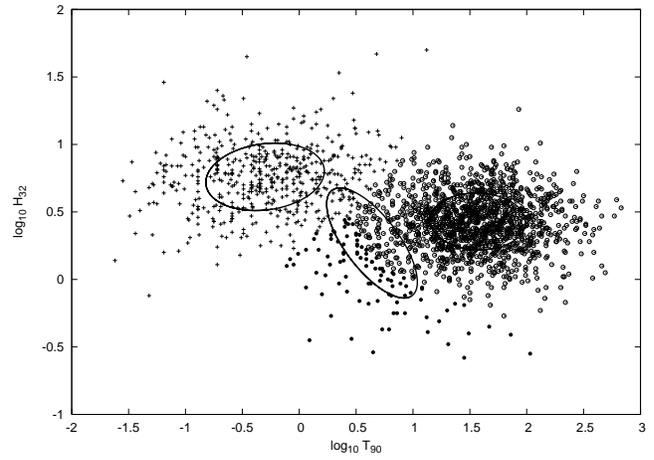}\\
  \caption{Distribution of $N = 1956$ GRBs in the $\{\log T_{90}; \log H_{32}\}$
plane. The $1\sigma$  ellipses of the three Gaussian distributions are
also shown, which were obtained in the ML procedure. The different
symbols (crosses, filled circles and open circles) mark bursts
belonging to the short, intermediate and long classes,
respectively.} \label{t90h32}
\end{figure}

According to Eq.(\ref{mprob}) each burst may belong to any of the
classes with a certain probability. In this sense one cannot
assign a given burst to a given class with certainty, but with a
given probability. This type of classification is called a "fuzzy"
classification \cite{mclaba88}. Although, any burst with a given
$[x,y]$ could be assigned to all classes with a certain
probability, one can select that $l$ at which the $I_l(x,y)$
indicator function reaches its maximum value. Figure \ref{t90h32}
shows the distribution of GRBs in the $\{\log T_{90}; \log
H_{32}\}$ plane, in which the classes obtained in this way are
marked by different symbols. The $1\sigma$ ellipses of the three
Gaussian distributions are also shown.

\subsection{Application of the fuzzy classification}

Inspecting Figure \ref{t90h32} one can recognize immediately that
the domain within the ellipse of the intermediate group is only
partly populated by GRBs belonging to this class according to the
classification procedure described above. The remaining part is
dominated by GRBs classified as short and, in particular, as long.
In other words, the ellipse of the third group contains an
essential amount of GRBs, which should belong either to the long
group or to the short group. Due to the 'fuzzy' classification
some probability was also assigned  to the other classes. Based on
the analytical expressions of the components, one can easily
calculate the contribution of any other groups within the ellipse
of a given class by summing the $I_l(x,y)$ values of different
$l$-s for the bursts lying in this particular region.

The reliability of the classification can be characterized by
counting the different classes of the GRBs lying within the
1$\sigma$ ellipse of a given Gaussian component. If the
classification were correct, only those GRBs would lie within the
ellipse of a given $l$ that have classes corresponding to this
component. Denoting by $n_l$ the number of GRBs within the ellipse
belonging to class $l$ one gets $n_1 = 218$, $n_2 = 174$, $n_3
= 514$. The rows of Table \ref{tab4} give the number of GRBs of
 all classes within the $1\sigma$
ellipses of the short, intermediate and long Gaussian components.
The first row shows that in the ellipse that defines the short
group, there are 218 GRBs. In accordance with the fuzzy
classification all have the highest probability assigning them to
the short group. Similarly, the third row shows that in the
ellipse, which defines the long group, there are 514 GRBs. All
these, in accordance with the fuzzy classification, have the
highest probability assigning them to the long group. But in the
second row, which defines the 174 GRBs in the ellipse defining the
intermediate group, only 47 bursts have the highest probability
assigning them to the intermediate group. A further 21 (106) GRBs
should belong to the short (long) class.

\begin{table}[h]  \centering
\caption{ Number of GRBs classified by the procedure described in
the text, within the $1\sigma$ ellipses of $l=1,2,3$ Gaussian
components. } \label{tab4}
 \begin{tabular}{ccccc}
\hline
           $l$   &  short &  interm. &  long  & Total \\
 \hline
             1.  &   218  &    -   &    -   &  \  218 \\
             2.  &    \ 21  &    47  &   106  &  \  174  \\
             3.  &     -  &     -  &   514  & \   514 \\ \hline
            Total  &   239  &     47 &   620  &    906 \\ \hline
  \end{tabular}
\end{table}

Table \ref{tab4} demonstrates  that the
classifications of the short and long GRBs are very reliable,
since they do not overlap the other two classes. This
means that GRBs within the ellipse of the first and third class
(first and third row in Table \ref{tab4}) were well classified as
short and long, respectively. In contrast, the ellipse of the
intermediate component (second row) contains a significant number
of members of the two other classes, in particular of the long
group. This is caused predominantly by the closeness of the most
numerous long class to the intermediate one.

There are $N - (n_1 + n_2 + n_3) = 1050$ GRBs scattered over a
much larger area outside the ellipses. In this region the Gaussian
components have low probabilities. The indicator function can
still have a large value, however, because there are small numbers
in both the nominator and denominator of the right-hand side of
Eq.(\ref{mprob}). Although the classification of these bursts is
formally correct, it is less reliable than those within the
ellipses.

We demonstrated the  robustness of classification by
comparing the results obtained from the $\{\log T_{90};
\log H_{321}\}$ and the  $\{\log T_{90};
\log H_{32}\}$ planes, respectively. A cross tabulation
between these two classifications is given in Table \ref{tab5}.
One may infer from this cross tabulation that the short and long
classes correspond within about 10\% to the respective groups
obtained from the other classification. Consequently, the
robustness of the short and long group is well established. On the
contrary, the population of the intermediate group is much poorer
in classifying in the $\{\log T_{90}; \log H_{32}\}$ plane
than in the other one. Table \ref{tab5} clearly shows that
classification ${\rm Class_{321}}$, except for one, contains all
GRBs assigned to the intermediate group by  ${\rm
Class_{32}}$.

Classification in the $\{\log T_{90}; \log H_{321}\}$
plane indicated 42  GRBs from the short and 89  ones from the long
groups, respectively. This high number of indicated bursts clearly
shows that a slight variation of the parameters of the Gaussian
distribution representing the intermediate group results in a
drastic change in the number of classified objects in this group.
However, comparing  the fraction of GRBs belonging to the
intermediate group according to Table \ref{tab2} and Table
\ref{tab32} one gets figures of 213 and 294, respectively.

If one assigned the burst to that group that had the maximum
membership probability a slight change in the parameters of the
corresponding Gaussian  distribution may move the GRB to an other group. On the
contrary, the fuzzy classification assigns membership probability
to all of the bursts. Hence, a small variation of the parameter gives a
small variation in the estimated number of bursts in the
intermediate group obtained by summing  the membership
probabilities of all GRBs in the sample.

\begin{table}[h]  \centering
\caption{ Cross tabulation  of GRBs classified in the
$\{\log T_{90}; \log H_{321}\}$ (${\rm Class_{321}}$) and
the $\{\log T_{90}; \log H_{32}\}$ (${\rm Class_{32}}$)
plane, respectively. \label{tab5}}
 \begin{tabular}{ccccc}
\hline &  &       ${\rm Class_{321}}$ &  & \  \\
\cline{2-4}
  ${\rm Class_{32}}$&      short &  interm.  &   long & Total \\
  \hline
 short             & 474 & 42 & 4 &  520 \\
 interm.    & - &  98 & 1 &  99 \\
 long       & - &  89 & 1247 &   1336 \\
\hline
 Total       &474 &229 & 1252  &  1955 \\
 \hline
 \end{tabular}
\end{table}

\subsection{Effect of observational bias on the classification}
\label{bias}
Performing several classification techniques on the whole BATSE
GRB sample indicates the intermediate group with a high
certainty. Hakkila et al. (2003) claimed the  structure of the BATSE sample
 identifying the intermediate group is due to a
special kind of observational bias. They pointed out that it is
reasonable to assume that observations of faint, long GRBs  detected
only the brightest part of the burst and a significant fraction
was buried in the background noise. It  also means an
underestimation of the true duration. As a consequence the faint
long bursts appear to be softer and shorter than in  reality.
This effect could produce the intermediate group in the sample.
Detailed study of this effect, however, has proven that the existence
of the intermediate group cannot be accounted for by it.

Hakkila et al. (2003) studied a further possibility which might be
the reason for the existence of the intermediate group. The
detection of the bursts proceeds on three timescales:  64 ms,
256 ms and 1024 ms. To record a GRB the count rate of the
peak-flux  has to exceed the detection threshold on at least one
of these time scales. A slow faint burst, which emitted the same
amount of energy as a shorter one, might be missed by the
observation since the peak-flux event on the longest 1024 ms
timescale was less that that of a faster one. Supposing a
Gaussian distribution for the logarithmic duration of the long
bursts, truncation of the slow faint GRBs results in a relative
overabundance of those that lie in the short duration wing.
Fitting this truncated distribution with Gaussian distributions
one obtains an additional component  accounting for the enhancement
at the short duration wing.

\begin{figure}
\includegraphics[width=6.1cm,angle=270]{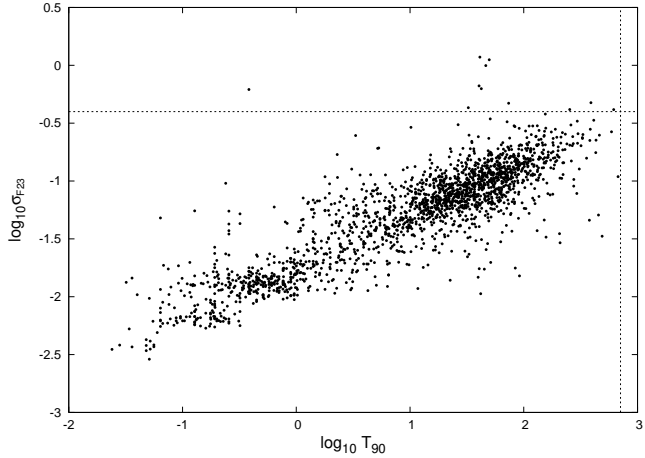}
 \caption{Relationship between $\log T_{90}$ and the
$log \sigma_{F32}$ error of the $F_{32}$ fluence. The vertical dashed line
indicates the duration of the longest bursts in the sample.
The horizontal dashed line marks the expected value of  $log \sigma_{F32}$
at the longest duration.} \label{s23}
\end{figure}

In the case of bursts where the duration is shorter than the time
scale of detection there is a one to one correspondence between
the peak-flux and  the total number of counts observed. As a
consequence, the fluence and the peak-flux on this time scale are
identical within a conversion factor. Let us suppose, in addition
to the 64, 256 and 1024 ms timescales, we have a further one which
is longer than the longest burst in the BATSE sample. Figure
\ref{s23} shows the relationship between the $\log T_{90}$
duration and the $\sigma_{F32}$ error of the $F_{32}= F_3 + F_2$
fluence. The horizontal dashed line marks the expected mean error
of the longest burst in the sample.

\begin{figure}
\includegraphics[width=6.1cm,angle=270]{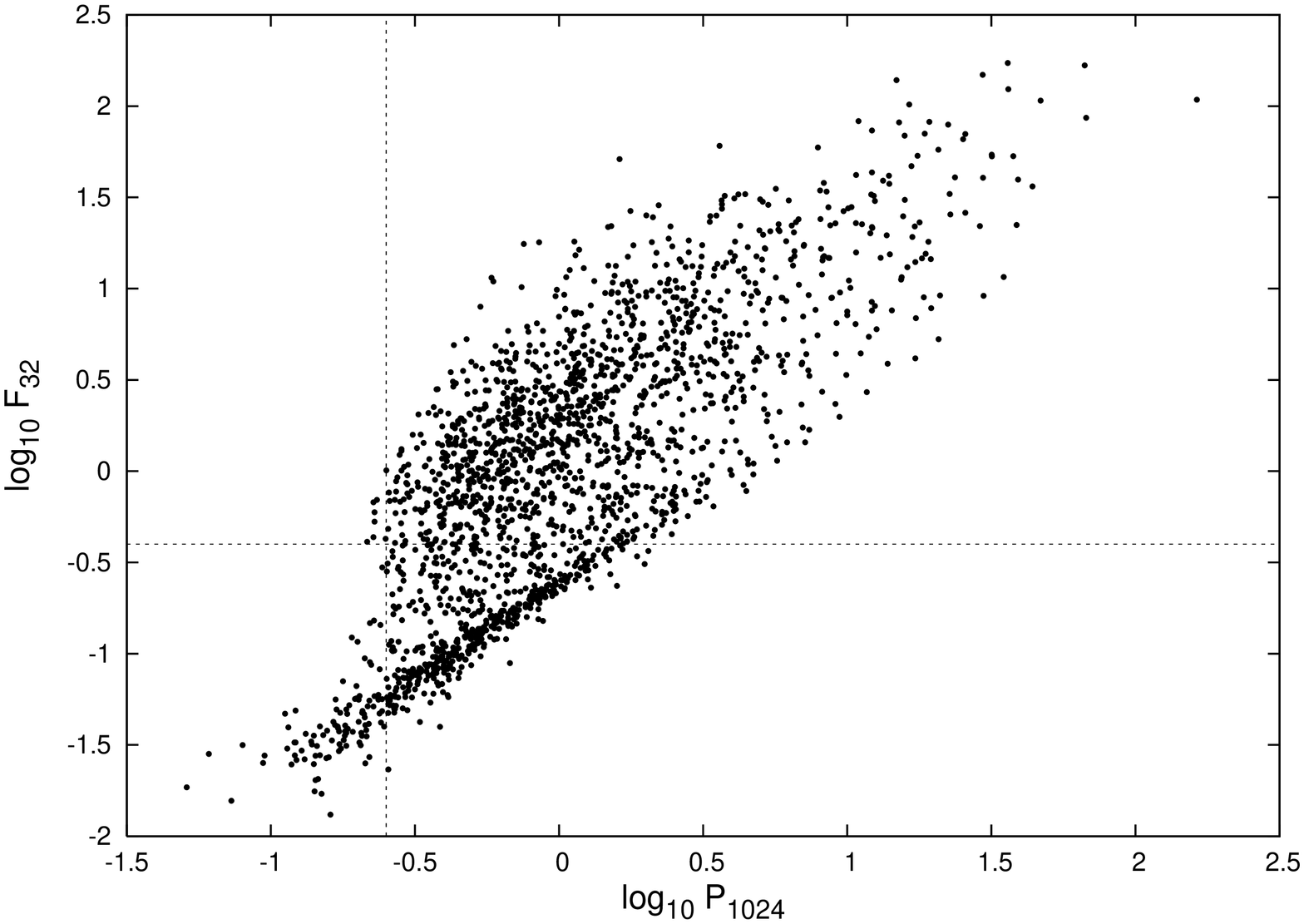}
\includegraphics[width=6.1cm,angle=270]{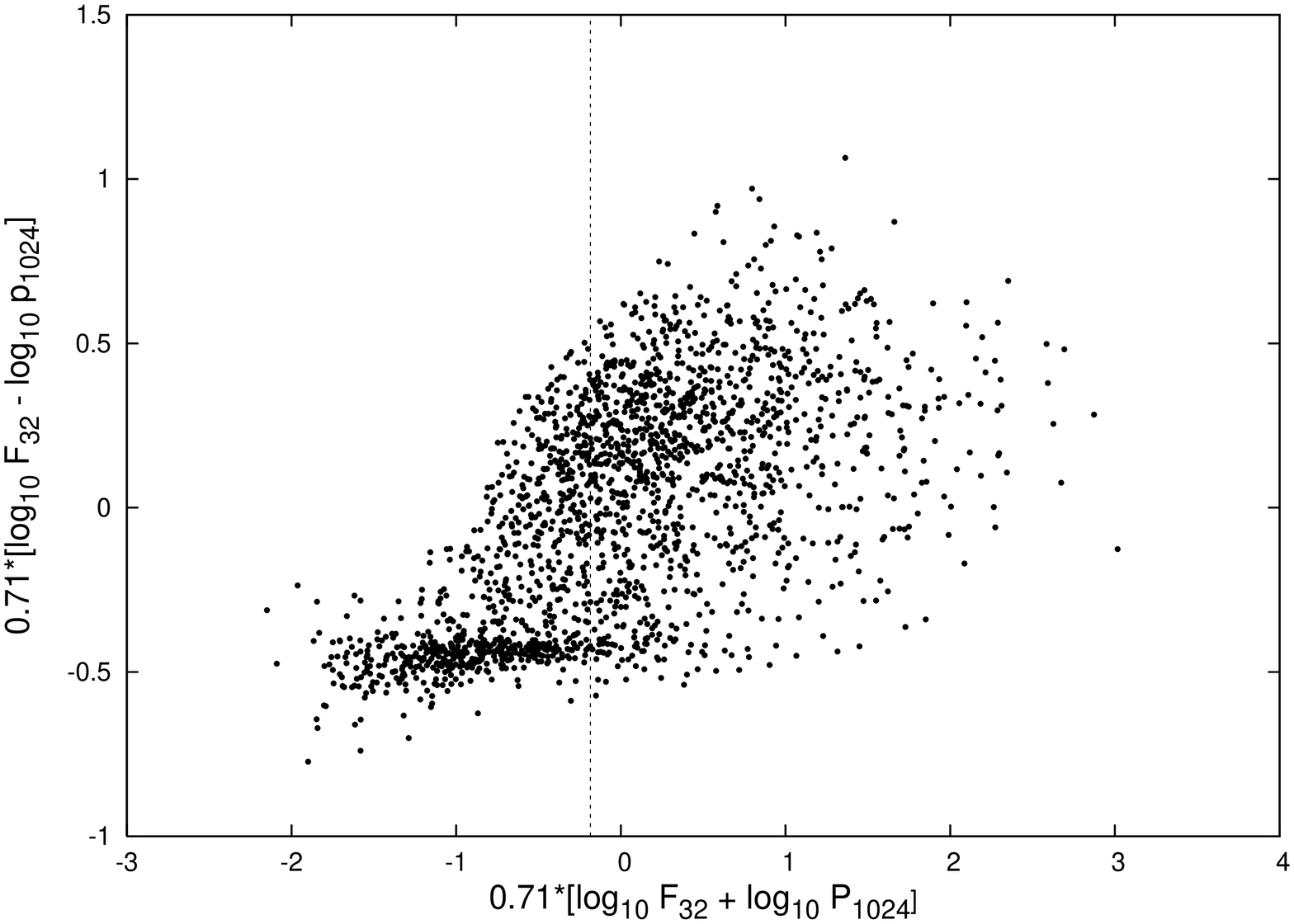}
 \caption{Distribution of the GRBs of the BATSE Current
Catalog in the $\{\log_{10} P_{1024}; \log_{10} F_{32}\}$ plane.
Vertical dashed line indicates the trigger level of the $P_{1024}$
peak flux and the horizontal one marks the expected value of $5.5
\times \sigma_{F32}$ at the longest duration (top panel).
Distribution of points in the top panel after $45^{\rm o}$
rotation (bottom panel). Vertical dashed line shows the limit of
completeness on the dual time scale defined in the text .}
\label{dusc}
\end{figure}

Let us take a hypothetic detection timescale as long as the
longest GRB in our sample. The detection is successful if the
fluence is greater than 5.5-times the noise level (this is the
usual BATSE trigger criterion). We marked this level  by the
horizontal line in the top panel of Fig. \ref{dusc}. A burst
fulfilling this criterion would be detected independently of
exceeding the trigger level on the other time scales. Following
 the idea of Hakkila et al. (2003) we introduced a dual
timescale from 1024 ms and the longest duration in the BATSE
Catalog  (800 $s$), in contrast to 10000 $s$ of Hakkila et al.
(2003). The difference between the two timescales may have an
 impact on the final classification.

Denoting with $P^{th}_{1024}$ and $F^{th}_{32}$ the detection
threshold on the 1024 ms and the hypothetical long timescale the
inequality $ P^{th}_{1024} + F^{th}_{32} < P_{1024} + F_{32}$
defines that part of the $\{\log_{10} P_{1024}; \log_{10}
F_{32}\}$ plane in which all the GRBs are detected.  
Replacing the inequality with the equality in the previous
relationship we obtain a line of $-45^{\rm o}$ slope which is the
boundary of completeness in this plane. Rotating the coordinates
by $45^{\rm o}$  the boundary of the area of completeness
becomes a vertical line as indicated in the bottom  panel of
Figure~\ref{dusc}.

Restricting ourself to the region of completeness in the 
$\{0.71[\log_{10} F_{32}+\log_{10} P_{1024}]; 0.71[\log_{10}
F_{32}-\log_{10}P_{1024}]\}$ plane (right of the vertical dashed
line in the bottom panel of Fig. \ref{dusc}) we repeated the
group-searching algorithm making use of this part of the BATSE
sample. Tables \ref{trun2}, \ref{trun3} and \ref{trun4} summarize
the results of the computation.

\begin{table}
\centering  \caption{Results of the EM algorithm on the truncated
sample.
  $k=2$ \ \ $L_{max}=1152$. (Details of the truncation are described
  in the text.)} \label{trun2}
\begin{tabular}{ccccccc}
\hline
 $l $ &  $p_l$ &  $a_x$ &  $a_y$ &     $\sigma_x$  &   $\sigma_y$  &  $r$ \\
\hline
 1  & 0.194  &   0.562 &  0.290  &    0.697 &   0.380 &  -0.574 \\
 2  & 0.806  &   1.631 &  0.178  &    0.391 &   0.238 &  -0.024 \\
\hline
\end{tabular}
\end{table}

\begin{table}
\centering  \caption{Results of the EM algorithm on the truncated
sample.
  $k=3$ \ \ $L_{max}=1172$} \label{trun3}
\begin{tabular}{ccccccc}
\hline
 $l $ &  $p_l$ &  $a_x$ &  $a_y$ &     $\sigma_x$  &   $\sigma_y$  &  $r$ \\
\hline
1 & 0.067 & 0.015  & 0.660  &  0.502 &  0.139 & -0.064 \\
2 & 0.130 & 0.932  & 0.046  &  0.652 &  0.322 & -0.391 \\
3 & 0.802 & 1.622  & 0.186  &  0.397 &  0.233 & -0.040 \\
\hline
\end{tabular}
\end{table}

\begin{table}
\centering  \caption{Results of the EM algorithm on the truncated
sample.
  $k=4$ \ \ $L_{max}=1175$} \label{trun4}
\begin{tabular}{ccccccc}
\hline
$l $ &  $p_l$ &  $a_x$ &  $a_y$ & $\sigma_x$ & $\sigma_y$  &  $r$ \\
\hline
1    &  0.063 & -0.049 &  0.663 &    0.486   & 0.138 &  0.002 \\
2    &  0.072 &  0.939 & -0.067 &    0.722   & 0.303 & -0.395 \\
3    &  0.041 &  0.592 &  0.271 &    0.259   & 0.291 & -0.566 \\
4    &  0.825 &  1.620 &  0.184 &    0.393   & 0.235 & -0.046 \\
\hline
\end{tabular}
\end{table}

The truncation procedure described above left 1229 GRBs in the
sample. Inspecting the results given in Tables \ref{trun2},
\ref{trun3} and \ref{trun4} one may infer that increasing the
number of Gaussian components from $k=2$ to $k=3$ yielded a
significant increase in the likelihood while from $k=3$
to $k=4$ did not. We conclude that even in this truncated
sample  some fraction (13\%) (it was 15\% in the
non-truncated case)  still appeared to belong to the
intermediate group. Comparing the $a_x$ parameters between Tables
\ref{tab32} and \ref{trun3} shows that the deviations are much
less than the corresponding $\sigma_x$ term.  It remains to show,
however, what fraction of the  intermediate
 GRBs in the whole sample was assigned to the same class in the truncated case.
In Table \ref{tab12} we made a cross tabulation between the
classification of the whole and the truncated sample in the $\{
\log T_{90}; \log H_{321}\}$ plane. This table  shows that out of
the 92 intermediate GRBs in the non-truncated sample 77 remain in
the same class in the truncated case  but 17 arrived from the
other two classes (12 from the short and 5 from the long group).

\begin{table}[h]  \centering
\caption{ Cross tabulation  of GRBs classified in the $\{\log
T_{90}; \log H_{321}\}$ plane in the truncated and non-truncated
cases, respectively.} \label{tab12}
 \begin{tabular}{ccccc}
\hline 
&   &      non-truncated  &  & \     \\
&  &       class${}_{321}$ &  & \  \\
\cline{2-4}
  truncated class${}_{321}$&      short &  interm.  &   long & Total \\
  \hline
 short             & 83 & 5 & 0 &  88 \\
 interm.    & 12 &  77 & 5 &  94 \\
 long       & 0 &  10 & 1037 &   1047 \\
\hline
 Total       &95 & 92 & 1042  &  1229 \\
 \hline
 \end{tabular}
\end{table}

\subsection{Caveats}
\label{cave}  The fuzzy classification assigned three
$\{p_1,p_2,p_3\}$ probabilities ($p_1+p_2+p_3=1$)  to each GRB in
the sample. Somewhat arbitrarily, we assigned the $k$ class to a
burst event where $p_k$, ($k=1,2,3$) was maximal. The fraction of
GRBs selected in this way could be different to $\sum
\limits^N_{i=1} p_k^i/N$, the expected percentage of class '$k$'
within the  whole population. The truncation we applied in Section
\ref{bias} affected the parameters of the best fitting Gaussians,
consequently $p_k$s, and it might move some bursts into another
class while others are added. The fraction of a class within
the whole population, however, could be more resistent than the
classification of individual objects. This fact implies that we
cannot classify the individual bursts with certainty in this way.

As the fuzzy classification required a functional form for a
suspected class to obtain membership probabilities, we assumed
Gaussian distributions. The results 
 reflect therefore a stochastic structure of the  sample rather
 than isolating a group of objects with  some distinct astrophysical properties.
 Consequently, it remained unclear at this stage whether the stochastic structure we
 uncovered by the EM algorithm really represents a  new class of
 GRBs. 

\section{Physical differences between the mathematical classes}

 In Section \ref{cave} we pointed out that the mathematical
deconvolution of the $p_l(x,y)$ joint probability density of the
observed quantities into Gaussian components does not necessarily
mean that the physics behind the classes obtained mathematically
is  different. It could well be possible that the true
functional form of the distributions is not exactly Gaussian and
that the algorithm of deconvolution  formally inserts a third one
only in order to get a satisfactory fit.  One needs detailed
investigations based
 on the physical (e.g. spectral) properties of the individual bursts to prove its
 astrophysical validity.

Recently Bal\'azs et al. (2003) found compelling evidence that
there is a significant difference between the short and long GRBs.
This might indicate that different types of engines are at work.
The relationship of long GRBs to the massive collapsing objects is
now also observationally well established \cite{me03}, and the
relation between the comoving and observed time scales is well
understood \cite{rype}. The short bursts can be identified as
originating from neutron star (or black hole) mergers
\cite{me01}. So the mathematical classification of GRBs into the
short and long classes - obtained  here (see Table \ref{tab1}
for $k=2$) and in Bal\'azs et al. (2003) - is also physically
justified.

An important question that must be answered in this context is
whether the intermediate group of GRBs, obtained in the previous
paragraph from the mathematical classification, really represents a
third type of burst physically different from both the short and
the long ones.

The classification into the short, intermediate and long classes
is based mainly on the duration of the burst. From
Table \ref{tab2} one may infer that these three classes differ
also in the hardnesses. The difference in the hardnesses between
the short and long group is well known \cite{kou}. According to
these data the intermediate GRBs are the softest among the three
classes. This different small mean hardness and also the
different average duration suggest that the intermediate group
should also be a different phenomenon, that is, both in hardness
and in duration  the third group  differs from the other two.
On the other hand, no correlation exists between the hardness and
the duration within the short and the long classes. More
precisely, no correlation exists for the long group and a  very
weak correlation exists for the short group (see Table
\ref{tab2}). Thus, these two quantities may be taken as
two independent variables, and the short and long groups are
different in {\it both} these independent variables.

 In contrast, there is a strong anticorrelation between the
hardness and the duration within the intermediate class. This is a
surprising, new result, and because the hardness and the duration
are {\it not} independent in the third group, one  may simply say
that only one significant physical quantity is responsible for
 the hardness and the duration within the intermediate
group. Consequently, the situation is quite different here,
because one needs two independent variables to describe the
remaining two other groups. This is a strong constraint in
modeling the third group.  Hence, the question of the true nature
of the  physics in the intermediate group remains open, and
obviously needs further detailed study.

\section{Conclusions}

Using the bivariate, duration-hardness fittings we obtained the
following results:

\begin{itemize}

 \item Increasing $k$ from 2 to 3 shows that the introduction of
the third group is real. This means that three groups of GRBs
should exist. This confirms the earlier results of several
authors.

 \item Increasing  $k$ from 3 to 4 shows that the introduction of
the fourth group is not needed. This means that {\it only} three
groups should exist. This result is in accordance with Mukherjee
et al. (1998). Discussion of the possible biases and also the use
of two different hardnesses do not change this conclusion.

 \item From the fitting procedures it follows that the duration and
the hardness are good quantities for the classification of GRBs.
Remarkably, the intermediate class is on average even
softer than the long group.

 \item We developed a method that makes it possible to define,
for any GRB, the probabilities determining its membership of a
given class. (The memberships are available by
internet \cite{ho05}.)

 \item 11\% - 15 \% of GRBs in the Current BATSE Catalog should belong to
the intermediate class.

 \item An unusual anti-correlation between the duration and hardness
might exist in the intermediate group. Hence, contrary to
the other two classes, here the duration and hardness
might not be independent variables, and hence the intermediate
class can be different from the other two classes
where the logarithmic hardness and duration are non-correlated
variables. Thus, further detailed analysis has to be carried out
to study this suspected behavior of the intermediate class.

\end{itemize}

All these considerations  mean that we answered
five questions of the six formulated in the Introduction.
The question "Is the intermediate group a physically different
phenomenon?" was not answered with satisfying  certainty,
and needs further analysis.

\begin{acknowledgements}
Thanks are due to D. L. Band, C.-I. Bj\"ornsson, J. T. Bonnell, L.
Borgonovo, I. Csabai, J. Hakkila, S. Larsson, P.
M\'esz\'aros, J. P. Norris, H. Spruit, G. Tusn\'ady and R. Vavrek for
valuable discussions. This study was supported by OTKA, grant No.
T034549 and T048870; by a Research Plan J13/98: 113200004 of The Czech
Ministry of Education, Youth and Sports (A.M.), and by a grant
from the Wenner-Gren Foundations (A.M.). A.M. wishes to
express his gratitude to 
 host institutes in Stockholm and Garching for their warm
hospitality. The detailed and
constructive remarks of  the anonymous referee significantly helped
to improve the scientific merit of our paper.
\end{acknowledgements}

\end{document}